\documentclass[12pt,preprint,natbib]{aastex}

\RequirePackage{lineno} 
\usepackage{longtable}
\usepackage{epstopdf}  
\usepackage{ifthen}
\usepackage{rotating}
\usepackage{subfigure} 
\usepackage{booktabs}

\def\rh{$r_{\mathrm{H}}$}

\begin{document}
\bibliographystyle{apj}

\setlength{\footskip}{0pt} 

\title{The Nucleus of Comet 10P/Tempel 2 in 2013 and Consequences Regarding Its Rotational State: 
Early Science from the Discovery Channel Telescope}

\author{David G. Schleicher\altaffilmark{1,2}, Matthew M. Knight\altaffilmark{2,3}, Stephen E. Levine\altaffilmark{2}}

\author{Submitted to {\it The Astronomical Journal} 2013 August 22; Accepted: 2013 September 10}

\altaffiltext{1}{Contacting author: dgs@lowell.edu.}
\altaffiltext{2}{Lowell Observatory, 1400 W. Mars Hill Rd, Flagstaff, AZ 86001, USA}
\altaffiltext{3}{Visiting scientist at The Johns Hopkins University Applied Physics Laboratory, 11100 Johns Hopkins Road, Laurel, Maryland 20723, USA}

\begin{singlespace}

\begin{abstract}
We present new lightcurve measurements of Comet 10P/Tempel 2 carried out 
with Lowell Observatory's Discovery Channel Telescope in early 2013 when 
the comet was at aphelion. These data represent some of the first science  
obtained with this new 4.3-m facility. With Tempel 2 having been observed 
to exhibit a small but ongoing spin-down in its rotation period for over 
two decades, our primary goals at this time were two-fold.  
First, to determine its current rotation period and compare it to that 
measured shortly after its most recent perihelion passage in 2010, 
and second, to disentangle the spin-down from synodic effects due to 
the solar day and the Earth's orbital motion and to determine the sense 
of rotation, i.e. prograde or retrograde. At our midpoint of 2013 Feb 24, 
the observed synodic period is 8.948$\pm$0.001 hr, exactly matching 
the predicted prograde rotation solution based on 2010 results, 
and yields a sidereal period 
of the identical value due to the solar and Earth synodic components 
just canceling out during the interval of the 2013 observations. The 
retrograde solution is ruled out because the associated sidereal 
periods in 2010 and 2013 are quite different even though we know that 
extremely little outgassing, needed to produce torques, occurred in 
this interval. 
With a definitive sense of rotation, the specific amounts of spin-down 
to the sidereal period could be assessed.  The nominal values imply that 
the rate of spin-down has decreased over time, consistent with the secular 
drop in water production since 1988. Our data also exhibited an 
unexpectedly small lightcurve amplitude which appears to be 
associated with viewing from a large, negative sub-Earth latitude, and 
a lightcurve shape deviating from a simple sinusoid
implying a highly irregularly shaped nucleus.
\end{abstract}


\section{INTRODUCTION}
Comet 10P/Tempel 2 is the second largest Jupiter-family comet in 
existence -- only 28P/Neujmin 1 is known to be larger -- and is 
comparable in size to 1P/Halley. Unlike Halley, however, Tempel 2 has a 
very small fraction of its surface that is still active ($<$1\%) and is 
therefore one of the most highly evolved comets. These two properties have 
made it possible to study its nucleus at both larger heliocentric distances 
\citep{jewitt88,mueller96} and at favorable perihelion 
passages \citep{jewitt89,ahearn89,wisniewski90,knight11a,knight12}. 
It also exhibits a strong asymmetry 
about perihelion in brightness \citep{sekanina79} and production rates due 
to the seasonal effects associated with the orientation of the rotation 
axis and a small source region near one pole that dominates the outgassing
\citep{knight12}. Of particular interest here is 
that Tempel 2 is one of the first comets to exhibit not just a change in 
its rotational period \citep{mueller96} but a progressive spin-down 
over multiple apparitions. Our lightcurve measurements obtained at the 
1999 apparition gave a period of 8.941$\pm$0.002 hr, about 32 s longer than 
the value of 8.932$\pm$0.001 hr two orbits earlier in 1988 based on a reanalysis 
of all available data \citep{knight11a}.  We followed this with further 
measurements in 2010, finding a period of 8.950$\pm$0.002 hr, thereby strongly 
implying the spin-down is caused by progressive torquing associated with 
the outgassing from the near-polar source region \citep{knight12}. 

The situation, however, is somewhat more complicated than it first appears. 
While the measured spin-down was the same between 1988 and 1999, and between 
1999 and 2010, the observations in the earlier apparitions were obtained 
several months prior to perihelion while the observations in 2010 were obtained 
between two and six months following perihelion. Thus a somewhat larger 
change in period was expected in the second pairing if the amount of torque 
remained the same at every perihelion passage. While our 2010 result might 
be explained by a secular decrease in outgassing that we observed during the 
past quarter century, it remained unclear what period Tempel 2 ultimately 
attained as it retreated from the Sun in 2011 and activity dropped off. 
Additionally, all of the 
reported values for its rotation period are synodic rather than sidereal. 
While some evidence has suggested that Tempel 2 is in prograde rotation
\citep{sekanina91}, a reanalysis of the same data gave conflicting results 
\citep{knight11a}. Since the spin-down of the nucleus adds a further 
complication to extracting the sense of rotation from varying synodic 
effects, we considered the direction of rotation as being unresolved, 
resulting in a pair of possible sidereal periods for each synodic result. 
For example, the 2010 synodic value of 8.950$\pm$0.002 hr corresponds 
to a prograde sidereal solution of 8.948$\pm$0.002 hr or a retrograde solution 
of 8.955$\pm$0.002 hr, where these offsets are asymmetric due to the Earth's 
orbital motion. 
Since the difference between synodic and sidereal 
varies with orbital position and observer geometries, this results in an 
added ambiguity to the specific amount of torque delivered each apparition. 
Fortunately, the sense of rotation can be resolved with additional measurements 
of Tempel 2 at a more distant location in its orbit when the synodic/sidereal 
difference is much smaller than near perihelion.

With these motivations, and with access to our new 4-meter class 
Discovery Channel Telescope (DCT), we decided to again measure Tempel 2's 
rotation period but this time near aphelion. The specific timing of the 
observations was based not only on the observing geometries but also 
required waiting for the comet to move sufficiently far away from the 
galactic plane. The initial night of planned observation also proved to be 
the first successful night of science operations for DCT, still early in its 
commissioning phase. We present here lightcurve data obtained over a 
3-month interval, and the resulting derived period along with the 
implications of this result.

\section{OBSERVATIONS AND REDUCTIONS}
\subsection{Instrumentation}
Lowell Observatory's Discovery Channel Telescope saw
first light in April 2012, less than 7 years after ground-breaking.
The telescope, with a 4.3-m primary, is located near Happy Jack, AZ, at an 
elevation of 2337 meters (cf. \citealt{levine12,bida12}). Site testing 
yielded a median seeing of 0.84 arcsec \citep{bida04} and delivered images 
already routinely exhibit FWHM of $\sim$0.8 arcsec when atmospheric conditions 
are good. The first instrument designed for DCT is the Large Monolithic 
Imager (LMI), which began service in September 2012. Containing a newly 
developed 6.1K$\times$6.1K CCD by e2v, this camera yields a 12.3 arcmin on a 
side field-of-view at the f/6.1 Ritchey-Chretien focus \citep{massey13}. 
Our observations were obtained using 2$\times$2 on-chip binning, resulting in a 
pixel scale of 0.240 arcsec, and we used a Kron-Cousins R band filter 
throughout.

\subsection{Observations and Reductions}
Successful and useful observations were acquired on two nights in January 
2013, one in March, and two in April; planned observations in 
February were snowed out while limited observations in May were interrupted 
by technical problems (including a regional power failure) making the 
remaining data useless for this project. Observational circumstances 
for each of the five useful nights are given in Table~\ref{table1}, including 
heliocentric distance ($r_\mathrm{H}$), geocentric distance ($\Delta$), and 
phase angle ($\theta$). The January and March runs each immediately 
followed large 
storms, resulting in quite poor and variable seeing ($\sim$1.0--2.4 arcsec), 
while conditions were better in April ($\sim$0.9--1.3 arcsec); image quality 
additionally degraded at high airmasses as we attempted to 
maximize the temporal coverage each night.
As expected during the very early science phase of a new telescope 
undergoing commissioning, a variety of small issues emerged and some 
affected the observations. In particular, pointing map problems in January 
required us to use the sidereal rate and manually guide on stars rather than 
track at Tempel 2's rate of motion. Four-minute integrations were used, 
resulting in 1.6 arcsec of trailing of the comet. 
We were able to use our preferred mode of tracking at the comet's rate 
of motion on the other runs, resulting in trailed comparison stars, 
though no tracking was actually required nor performed in April because the comet 
was within days of its stationary point. With a shorter exposure of 150 s 
and the slower rate of motion of the comet as it approached its stationary 
point, stellar trailing was quite small, 0.6 arcsec, in March while the comet 
trailed by $<$0.1 arcsec in April.


The choice of the aperture size for photometric extractions was, as usual, 
a compromise between a variety of factors, including trailing and seeing as 
just described, along with the contrast of the object to the sky, and how 
crowded the field was. The faintness of the comet and the occasional 
close passage of field stars argued for a small aperture while the other 
factors pushed for larger sizes, especially in January. Fortunately, 
as discussed below, no coma was detected and so no removal was necessary. 
We ultimately chose to minimize the resulting scatter in Tempel 2's 
lightcurve each night, which resulted in aperture radii of 11 pixels 
(2.6 arcsec) in January, 8 pixels (1.9 arcsec) in March, and 6 pixels 
(1.4 arcsec) in April; sky measurements were obtained from the median value 
in an annulus between 30 and 60 pixels (7.2--14.4 arcsec).
The effects of changing seeing were directly compensated for by using 
identical apertures for on-frame comparison stars (see below), while the 
differential effects associated with trailing were minimized with these 
relatively large extraction apertures.

Calibration was performed using standard techniques with the Interactive 
Data Language software package for removing bias and applying median twilight 
flat fields. An extinction coefficient of 0.098 mag per unit airmass was 
determined on Jan 13, and was applied to all of each nights' data as a first-order 
correction for airmass. A minimum of seven comparison stars were used each night; 
the relatively slow motion of the comet at aphelion coupled with the large 
field of view of LMI meant these stars remained in the field throughout 
the night. Remaining variations from frame to frame due to cirrus, 
changing seeing, and second-order airmass effects were compensated for 
by determining and applying the median value of the deviations of all 
the comparison stars. In practice, though the resulting standard 
deviations for the comparison stars were quite small, the scatter for the 
much fainter comet became unacceptable when this correction was 
larger than 0.5 mag, and we therefore excluded data above this threshold.
We additionally excluded frames for which the comet was adversely affected 
by nearby stars, cosmic rays, etc. The uncertainties on each resulting 
measurement of Tempel 2 were based on the photometric uncertainties combined 
with the standard deviations of the comparison star corrections.

Absolute flux calibrations were applied by using SDSS r filter catalog 
values of each of our comparison stars. The average nightly uncertainty
in the absolute flux calibrations was 0.06 mag, which encompasses both 
the uncertainty in the catalog magnitudes themselves and color effects 
in the extinction correction. The resulting nightly instrumental 
correction varied by as much as 0.3 mag because several nights had no 
interval when the sky was photometric. While Landolt fields were observed 
on the few photometric occasions, these measurements were too sparse to 
significantly improve the absolute calibration but did confirm that our 
calibration was good to better than 0.1 mag. Assuming a solar color,
we applied an average offset of 0.22 mag to convert from SDSS r 
to Cousins R\footnote{http://www.sdss.org/dr5/algorithms/sdssUBVRITransform.html}. 
Final lightcurve measurements and uncertainties of all 422 useful data 
points (as defined above) are listed in Table~\ref{table2}, as well as 
the observed mid-times for each image. 


\section{LIGHTCURVE RESULTS}
\label{sec:lightcurve_results}

\subsection{Lightcurves}
\begin{figure}
  \centering
  \includegraphics[width=88mm]{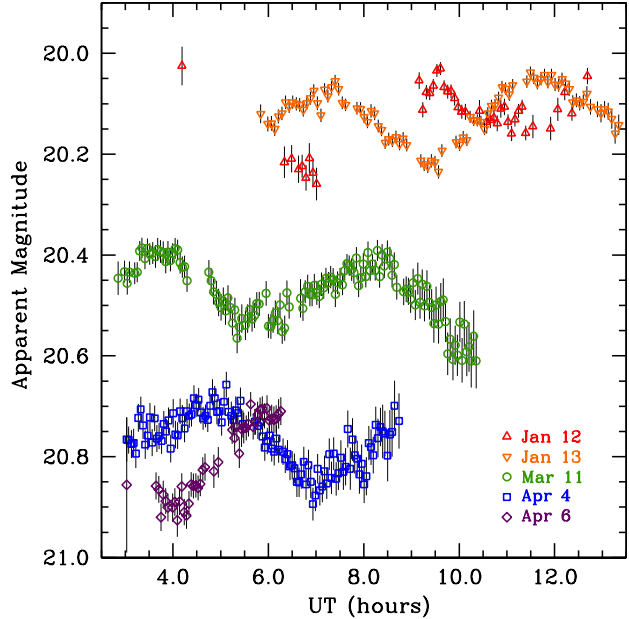}
  \caption[Unphased data]{Reduced R-band magnitudes (m$_R$) for our data plotted as a function of UT on each night. The magnitudes have had absolute calibrations, extinction corrections, and comparison star corrections applied and are given in Table~\ref{table2}. The symbols are defined in the legend. Note that while $r_\mathrm{H}$ remained approximately constant during the observations, {\it $\Delta$} and $\theta$ increased, causing the brightness to steadily decrease from month to month.}
  \label{fig1}
\end{figure}

The reduced lightcurves are displayed in Figure~\ref{fig1} as a function of time 
for each of the five nights. Immediately evident is the large decrease 
in brightness during the apparition that will be shown later to be simply 
due to the combination of increasing distance and increasing phase angle. 
Gaps in the data on Jan 12 and the apparent late start on Jan 13 are due 
to clouds and/or the comet passing too close to a star. 
Series of images having the best effective seeing from Jan 13 and again 
from Apr 6 were stacked to search for evidence of a coma; none was 
detectable on either night. Comparison of the wings of the comet's radial 
profile to stellar radial profiles on Apr 4 and Apr 6 (when the 
non-sidereal motion was negligible) suggests than any non-stellar 
component had an upper limit of 5\% of the total cometary flux. Therefore 
no coma removal was required (or possible) from these observations, 
unlike observations near perihelion. 


To remove the observed secular trend, we next performed a standard asteroidal 
normalization for these nuclear magnitudes to absolute values by correcting 
for heliocentric and geocentric distances and to zero phase angle, i.e.\ 
from the reduced magnitude, $m_\mathrm{R}$, to $m_\mathrm{R}$(1,1,0). 
Our first attempts at normalization used the same linear phase function 
that we applied in our previous papers \citep{knight11a,knight12}. This 
failed, however, to yield consistent brightnesses among the three observing 
runs no matter what phase coefficient was used. In retrospect, this was 
not surprising due to the much smaller phase angles involved in 2013 and the 
typical asteroidal non-linear opposition effect. Rather than empirically 
determining the needed curvature, we decided to use a model phase curve that 
best matched C-type asteroids, those with physical characteristics such as 
albedo most similar to comet nuclei. Fitting and interpolating the values from 
Table IV of \citet{bowell79}, we computed phase adjustments 
($\Delta$$m_\theta$) for 
each night as given in Table~\ref{table1}. The resulting 
$m_\mathrm{R}$(1,1,0) values at lightcurve maxima were in 
remarkable agreement, indicating that the phase function for Tempel 2's 
nucleus is indeed well-represented by C-type asteroids, and no further 
adjustments were applied.

\subsection{Rotation Period}


We continue to use our interactive period search routine for period 
determinations. As described in more detail in \citet{knight11a},
phase plots are instantaneously updated as we systematically step through 
possible periods and examine by eye the tightness of resulting lightcurves. 
Data are color-coded by date, making it easy to see small offsets in either 
brightness or phase throughout the phased curve. This is particularly 
useful when the intrinsic scatter due to photometric uncertainties is 
larger than usual, as we have here; a slightly incorrect period will 
only have a very slight increase in scatter and most period search 
algorithms will not ``see'' that one day is offset from another. In fact, 
we also used the phase dispersion minimization method \citep{stellingwerf78}
and, while this gave the same result, its uncertainty is much greater 
than what can be easily estimated from the color-coded phase plots, i.e.\ 
when clear shifts between observing runs is evident. Note that standard 
routines can also misinterpret changes in the shape of the lightcurve, 
another factor here.

\begin{figure}
  \centering
  \includegraphics[width=88mm]{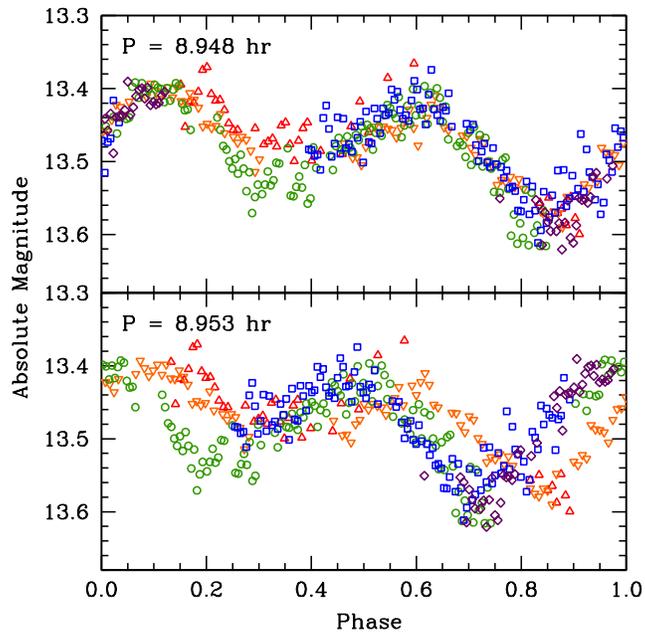}
  \caption[Data phased to 2010 prograde and retrograde solutions]{The lightcurve data phased to 8.948 hr (top) and 8.953 hr (bottom). Symbols are as given in Figure~\ref{fig1} and the magnitudes are normalized to (1,1,0) as described in the text. The 8.948 hr period is the synodic period expected in 2013 based on the 2010 prograde sidereal solution found in \citet{knight12} while the 8.953 hr period is the synodic period expected in 2013 based on the 2010 retrograde sidereal solution. The 8.953 hr period is clearly incompatible with the 2013 data, thus definitively ruling out retrograde rotation.}
  \label{fig2}
  \label{lastfig}
\end{figure}

Our overall best observed period solution was 8.948$\pm$0.001 hr, and 
the corresponding phase plot is shown in the top panel of Figure~\ref{fig2}.  
Since there is no obvious preferred zero point for phasing, we choose 
2013 Jan 0.0 as this placed zero phase away from interesting features 
and the $\Delta$$T$ values correspond to the day of the year; prior to 
phasing, the $\Delta$$T$ values were first corrected for light travel time 
from the observed mid-times in Table~\ref{table1}. 
As with previous apparitions, we have a double-peaked light curve where 
the two maxima are nearly identical in brightness but the minima are 
quite different, with the deeper minimum nearly ``V''-shaped while 
the smaller minimum varied in depth and shape with changing viewing 
geometry. Unfortunately, our phase coverage is incomplete in April and 
the shallower minimum is not sampled, so we cannot determine if the 
change in depth from January to March continued. Also evident is a 
$\sim$0.02 mag shift of much (but not all) of the Jan 12 lightcurve from 
that of a day later on January 13, for which we have been unable to 
identify a cause. While this has no effect on the the period solution, 
it alters the perception of the shape for the first half of the lightcurve. 


We also phased subsets of the data to look for evidence of a change in 
the observed period associated with the changing viewing geometries. 
As expected, phasing Jan/Apr yielded the same result as the entire 
data set, while a minor change was evident for the other two pairings -- 
Jan/Mar gave a slightly smaller value (8.947$\pm$0.001 hr) and Mar/Apr gave 
a slightly larger value (8.949$\pm$0.002 hr) with the higher uncertainty in the 
latter value directly caused by the shorter interval. 
Within the uncertainties, this increase matches the predicted change 
of $+$0.001 hr, caused simply by the Earth's orbital motion between our 
earliest midpoint pairing and the latest; the solar component to the 
synodic period near aphelion is both small and unchanging. 
These calculations all use the pole orientation from \citet{knight12}
of R.A.\ = 162$^{\circ}$ and Dec.\ = $+$58$^{\circ}$ (and the diametrically 
opposite solution for retrograde rotation), but similar pole positions, 
such as from \citet{sekanina91}, would yield the same results.

As discussed in the Introduction, one goal of these observations was to 
uniquely discriminate between prograde and retrograde rotation for 
Tempel 2's nucleus.  The sense of rotation always determines the sign of the 
solar component of the synodic period while the Earth's component is 
unaffected by the direction of rotation. 
At the midpoint of our ensemble of measurements for 2013, 
the solar and Earth components are coincidentally nearly identical and 
in the prograde case have opposite signs that cancel out, resulting in 
the sidereal period being the same as the synodic (8.948$\pm$0.001 hr), 
while in the retrograde case they compound yielding a sidereal value 
of 8.950$\pm$0.001 hr.

To determine which solution is correct, we next re-examine the solutions 
from 2010. Our overall observed (synodic) value was 8.950$\pm$0.002 hr with a 
corresponding midpoint time of 2010 Oct 25 or 112 days following perihelion 
\citep{knight12}. Based on our water production rates throughout the 
apparition, over 95\% of the total outgassing had taken place by this date. 
With an observed change in period of about $+$0.004 hr per perihelion passage, 
we concluded that further torquing late in the 2010 apparition should be 
negligible and that the sidereal period in 2013 would therefore be unchanged 
from the late 2010 result. The corresponding sidereal values at the 
2010 midpoint are 8.948 hr (prograde) and 8.955 hr (retrograde). 
Note that these values are slightly different from those listed in
\citet{knight12} which erroneously only included the solar component 
but not the Earth's component.
With the expectation that the sidereal value should be the same in early 
2013 as in late 2010, and the possible sidereal values in 2013 being 
8.948$\pm$0.001 hr (prograde) and 8.950$\pm$0.001 hr (retrograde), we conclude 
that Tempel 2 {\it must} be in prograde rotation, with a sidereal period 
in late 2010 and in early 2013 of 8.948 hr.

As further evidence that Tempel 2 is in a prograde rotation, we also 
examined the retrograde scenario. If it was the retrograde case, then as 
already stated the sidereal period in 2010 would have been 8.955 hr. 
Since the difference between the retrograde synodic and sidereal periods in 
2013 would be $-$0.002 hr, we should have measured a period of 8.953 hr. 
Phasing with this value is shown in the bottom panel of Figure~\ref{fig2} and it 
is clear that this solution does not yield a viable lightcurve, therefore 
the comet must instead be in prograde rotation.

Looking back at the earlier apparitions requires disentangling the synodic 
effects from the actual changes in period due to torquing. 
An early attempt to derive the sidereal period from the changing synodic 
period during 1988 \citep{sekanina91} suffered from several problems, including 
the claim of a smoothly decreasing period which could not be reproduced 
during a detailed reanalysis \citep{knight11a} and our new determination 
that his synodic/sidereal modeling only included the Earth's motion but 
neglected the dominant solar component. Having now definitively 
determined the sense of rotation as prograde, we can reexamine 
prior apparitions. The computed sidereal periods for each epoch are 
given in Table~\ref{table3}; for 
completeness and to illustrate the asymmetries of the pro- and retrograde 
solutions with respect to the observed synodic periods, we also tabulate 
the retrograde values. Again, these values differ somewhat from the 
values given in \citet{knight12} since the Earth component is now 
included along with the solar component, thereby causing the asymmetry. 
Our conclusions regarding a decrease in the amount of spin-down remain 
unchanged from \citet{knight12}, with sidereal periods before perihelia 
in 1988 of 8.931 hr and in 1999 of 8.939 hr, for an average spin-down by 
0.004 hr per apparition. The same rate of spin-down would have predicted 
a sidereal period of 8.951 hr in late 2010, a larger change than we observe. 
We continue to think the most likely cause of this is the decrease in total 
water production from 1988 to 1999 and to 2010, with lower production rates 
providing a smaller amount of torquing. However, within the uncertainties 
of the period determinations, the data are also consistent with no change 
in the rate of spin-down.
The agreement between the late 1994 period with early 1999 also remains, 
with both yielding a sidereal period of 8.939 hr during an interval 
for which there was not an intervening perihelion passage. 



\subsection{Nucleus Cross-Section}
Our 2013 lightcurves of Tempel 2 revealed two unexpected findings.  
First, the amplitude within the lightcurve was smaller than we had 
assumed it would be and, second, the peak brightness was higher 
than predicted from earlier apparitions even after compensating for 
a non-linear phase function. After considering a variety of possibilities, 
such as removing too much or too little coma for differing apparitions 
or systematic effects when different filters were used, 
we think we have arrived at a self-consistent explanation. 
Our expectations for 2013 were based on assuming that Tempel 2 is a prolate 
ellipsoid with similar dimensions for the small and intermediate axes while 
the long axis is about 2.1$\times$ greater. This is based on the 1988 thermal IR 
lightcurve from \citet{ahearn89} and that the sub-Earth latitude then 
varied between $-$1$^\circ$ and $-$9$^\circ$, i.e.\ the comet was viewed in 
1988 nearly equator-on using our preferred pole solution. 
In early 2013, the sub-Earth latitude was $-$47$^\circ$, yielding a predicted 
amplitude of 0.4 mag, significantly greater than the observed value of 
$\sim$0.2 mag. If, however, the intermediate axis is actually intermediate 
in length rather than the same as the small axis, then the total cross-section 
will appear to increase as one views from more pole-on and the amplitude 
will become even smaller than the simple function of the cosine of the 
sub-Earth latitude (the intermediate axis begins to dominate over the 
small axis). 
Note that in this tri-axial case, it is the ratio of the long axis to the 
intermediate axis that determines the amplitude for the equator-on view. 
Therefore, to explain the 1988 amplitude, the intermediate axis 
remains the same as originally assumed for 
the prolate case, and it is the short axis which must be even shorter. 
For our observed amplitude in 2013, this requires the short axis to be 
less than one-third that of long axis, while the intermediate axis remains 
at about one-half of the long axis. This implies an even more elongated 
nucleus than the tri-axial solution with ratios of $0.43:0.60:1.0$ proposed 
by \citet{sekanina91}. In addition to explaining the observed amplitude, 
our tri-axial solution also naturally explains the increased brightness 
at lightcurve maxima in 2013 as being due to the increased total 
cross-section relative to the equator-on view. 

We strongly suspect, however, that the shape of the nucleus significantly 
departs from a tri-axial ellipsoid for several reasons. The lightcurve 
shape is not sinusoidal, but rather has one ''V''-shaped minimum and one 
rounded minimum. Also, the shapes and depths of the minima change 
rapidly in 2013 with only a small change in viewing geometries even though 
the sub-Earth and sub-solar latitudes did not vary. Finally, as discussed 
below, the amplitudes are quite different when the comet was viewed from 
high positive latitudes as compared to high negative latitudes. Thus, 
we conclude that the nucleus must have large-scale protuberances. 

To first-order, the tri-axial ellipsoid explains most of the brightness and 
amplitudes measured at the other apparitions. Most similar to 2013 were 
the circumstance in early 1987 with a sub-Earth latitude of $-$52$^\circ$, 
and where Jewitt \& Meech (1988) measured 
an amplitude of $\sim$0.3 mag and a peak magnitude essentially identical 
to our 2013 value once adjustments are made for filter and phase angle. 
The 1999 apparition was nearly identical to that of 1988, and the 
lightcurve characteristics are also essentially the same. The peak 
brightness in 1994 \citep{mueller96} is also consistent with 
1988 and 1999 but the amplitude is somewhat smaller (by $\sim$0.1--0.2 mag) 
with no clear explanation since the sub-Earth latitude was close 
to the equator. Most different were our results from 2010 \citep{knight12}, 
where the peak brightness is $\sim$0.2--0.3 mag fainter while the 
amplitude is larger than expected for a sub-Earth latitude near $+$40$^\circ$. 
While an over removal of coma might partially explain both aspects, 
our methodology was the same used for the 1999 apparition which has 
no such issues, leading us to 
suspect that Tempel 2's actual shape characteristics are another 
significant factor; this is supported by the fact that only in 2010 
did the sub-Earth latitude have a high positive value ($+$40$^\circ$) while 
all other apparitions were either near the equator (1988, 1994, 1999) 
or at high negative values (1987, 2013).

\section{DISCUSSION AND SUMMARY}
\label{sec:discussion}

The observations reported here represent some of the first science 
collected with Lowell Observatory's new Discovery Channel Telescope. 
Despite a few problems associated with a facility still in the early 
stages of commissioning, we were able to easily fulfill our 
major science objectives regarding Comet Tempel 2's rotational state, 
in spite of a smaller lightcurve amplitude than expected. We obtained 
a precise measurement of its current rotation period that, when 
combined with our 2010 measurements, yielded a definitive determination 
of its sense of rotation. This result in turn allowed us to determine 
the correct sidereal period associated with observed synodic values 
at each prior apparition. For Lowell Observatory and its partners, a new 
era has begun in which projects requiring time-intensive or long-duration 
observations can be accomplished for targets much fainter than have 
ever been possible. In the specific case of comet nuclei, we can now 
measure lightcurves of comet nuclei far from perihelion.

The measured periodicity in a nucleus lightcurve depends on 
the changing cross-section both as illuminated by the Sun and as seen 
from the Earth, and in turn depends on the solar day as well as the 
Earth's motion with respect to the comet and the Sun. Disentangling 
these effects from the change in the physical rotation due to torquing 
had been problematic at best. However, the relative importance of 
the solar component and Earth's motion component are quite different 
far from the Sun as compared to near perihelion. Because of this, 
we planned and executed these new observations. The resulting synodic 
period, 8.948$\pm$0.001 hr, matched that expected for the prograde 
scenario from 2010 while the nominal period associated with the 
retrograde solution is clearly ruled out. We therefore conclude that 
Tempel 2's rotation is prograde with respect to both the ecliptic and 
to its orbital plane. The sidereal period, 8.948$\pm$0.001 hr, 
is coincidentally identical to the synodic value as the solar and Earth 
components of the synodic period just cancel out during the interval 
of our 2013 observations. The new data also confirm that 
no additional spin-down took place following our late 2010 measurements, 
consistent with the water production curve showing a near-cessation 
of activity by that time. This lack of a change in the sidereal period 
very late in the 2010 apparition thereby implies that the smaller change 
in the period per perihelion passage between 1999 and 2010 as compared 
to 1988 to 1999 is real (although a constant rate of spin-down cannot 
be excluded within the uncertainties). Note that the inferred changes 
match those given by \citet{knight12} for the prograde case. 
Thus our hypothesis that a decrease in torquing 
is associated with the secular decrease in water production from 1988 
to 2010 is probably correct.  We note, however, 
that there is no strong evidence for a long-term decrease in activity 
over the past century, and the secular drop over the past two decades 
may reverse as the surface is eroded and varying proportions of ice 
are exposed from successive perihelion passages.

Similar to prior apparitions, the light curve is double-peaked with 
near-equal maxima but quite differently shaped minima. The shape and 
depth of one of the minima also changed over the three-month interval, 
another trait observed in the past. Both characteristics directly 
indicate that the shape of Tempel 2's nucleus is not a simple prolate 
ellipsoid or even a triaxial ellipsoid 
but must instead have large-scale protuberances to cause 
such clear changes in the lightcurve from such a small change in 
viewing geometries ($<$20$^\circ$) and essentially no change in sub-Earth 
and sub-solar latitudes.  The unexpectedly small amplitude we 
measured in the lightcurve ($\sim$0.2 mag) appears to also be due to 
peculiarities of the nucleus shape -- our value is in good agreement 
with the only other data obtained at a similar sub-Earth latitude 
\citep{jewitt88}, while data taken from near-equator on 
or from the other hemisphere all exhibit a much larger amplitude. 
The data further suggest that the length of the short axis 
is substantially shorter than that of the intermediate axis, though the shape 
is probably not as extreme as that of Comet 103P/Hartley 2 as imaged 
from EPOXI \citep{ahearn11a}. Thus a more complete story of the properties 
of Tempel 2's nucleus is beginning to emerge. While determining the detailed 
shape must continue to await an upclose view -- Tempel 2 was the proposed 
target of the planned Comet Rendezvous Asteroid Flyby mission more than 
two decades ago -- additional numerical modeling and new investigations 
from afar can continue to reveal clues to its physical structure and 
evolving behavior.

\section*{ACKNOWLEDGMENTS}
We thank T. Farnham for calculations of the offsets between the prograde 
and retrograde synodic periods with respect to a given sidereal period. 
We gratefully acknowledge the assistance of M. Fendrock  
with the first observing run and preliminary analyses, and 
A. Venetiou, M. Sweaton, J. Sanborn, R. Winner, and S. Strosahl for their 
successful operations during the early commissioning phase of the DCT, 
thereby making these observations possible. We also thank R. Millis, 
W. L. Putnam, and J. Hendricks for their vision and support which 
allowed the DCT to become a reality, and all members of the DCT and 
LMI teams.

These results made use of Lowell Observatory's Discovery Channel Telescope, 
supported by Lowell, Discovery Communications, Boston University, 
the University of Maryland, and the University of Toledo. 
M.M.K. is grateful for office space provided by the University of 
Maryland Department of Astronomy and Johns Hopkins University Applied 
Physics Laboratory while working on this project.
The LMI instrument was funded by the National Science Foundation via grant 
AST-1005313.
This research has been supported by NASA's Planetary Astronomy Program 
(Grant NNX09AB51G).


\label{lastpage}

\end{singlespace}


\renewcommand{\baselinestretch}{0.78}
\renewcommand{\arraystretch}{1.0}


\begin{deluxetable}{lcccccccccccccc}  
\tabletypesize{\scriptsize}
\tablecolumns{15}
\tablewidth{0pt} 
\setlength{\tabcolsep}{0.05in}
\tablecaption{Summary of Tempel 2 observational circumstances in 2013.\,\tablenotemark{a}}
\tablehead{   
  \multicolumn{1}{l}{UT Date}&
  \colhead{UT}&
  \colhead{$\Delta$$T$\,\tablenotemark{b}}&
  \colhead{\rh}&
  \colhead{$\Delta$}&
  \colhead{$\theta$}&
  \colhead{$\Delta$$m_\theta$\,\tablenotemark{c}}&
  \colhead{Ecl. Long.}&
  \colhead{Ecl. Long.}&
  \colhead{Conditions}\\
  \colhead{}&
  \colhead{}&
  \colhead{(day)}&
  \colhead{(AU)}&
  \colhead{(AU)}&
  \colhead{($^\circ$)}&
  \colhead{(mag)}&
  \colhead{Earth ($^\circ$)\,\tablenotemark{d}}&
  \colhead{Sun ($^\circ$)\,\tablenotemark{e}}&
  \colhead{}
}
\startdata
Jan 12&3:34--13:29&$+$12.4&4.697&3.770&\phantom{1}4.5&0.42&203.8&309.3&Clouds\\
Jan 13&3:35--13:27&$+$13.4&4.698&3.765&\phantom{1}4.3&0.41&204.1&309.4&Intermittent clouds\\
Mar 11&2:51--10:38&$+$70.3&4.709&3.970&\phantom{1}8.8&0.64&225.4&313.1&Cirrus\\
Apr 4&3:01--\phantom{1}8:52&$+$94.2&4.707&4.293&11.6&0.76&234.6&314.6&Cirrus\\
Apr 6&2:56--\phantom{1}8:45&$+$96.2&4.706&4.323&11.8&0.76&235.5&314.8&Clouds\\
\hline
\enddata
\tablenotetext{a} {All parameters were taken at the midpoint of each night's observations.}
\tablenotetext{b}{Time since 2013 Jan 0.0.}
\tablenotetext{c}{Magnitude correction to normalize to $\theta$ = 0$^\circ$.}
\tablenotetext{d}{Ecliptic longitude of the Earth as seen from the comet.}
\tablenotetext{e}{Ecliptic longitude of the Sun as seen from the comet.}
\label{table1}
\end{deluxetable}


\renewcommand{\baselinestretch}{0.78}
\renewcommand{\arraystretch}{1.0}


\begin{deluxetable}{lccccclccccclccccclccc}
\tabletypesize{\scriptsize}
\tablecolumns{22}
\tablewidth{0pt} 
\setlength{\tabcolsep}{0.03in}
\tablecaption{Photometry of Comet Tempel 2 in 2013}
\tablehead{   
  \multicolumn{1}{l}{Date\tablenotemark{a}}&
  \colhead{UT\tablenotemark{b}}&
  \colhead{m$_R$\tablenotemark{c}}&
  \colhead{$\sigma_{m_{R}}$\tablenotemark{d}}&
  \colhead{}&
  \colhead{}&
  \colhead{Date\tablenotemark{a}}&
  \colhead{UT\tablenotemark{b}}&
  \colhead{m$_R$\tablenotemark{c}}&
  \colhead{$\sigma_{m_{R}}$\tablenotemark{d}}&
  \colhead{}&
  \colhead{}&
  \colhead{Date\tablenotemark{a}}&
  \colhead{UT\tablenotemark{b}}&
  \colhead{m$_R$\tablenotemark{c}}&
  \colhead{$\sigma_{m_{R}}$\tablenotemark{d}}&
  \colhead{}&
  \colhead{}&
  \colhead{Date\tablenotemark{a}}&
  \colhead{UT\tablenotemark{b}}&
  \colhead{m$_R$\tablenotemark{c}}&
  \colhead{$\sigma_{m_{R}}$\tablenotemark{d}}
}
\startdata


Jan 12&\phantom{1}4.188&20.025&0.039&&&Jan 13&\phantom{1}7.857&20.109&0.013&&&Mar 11&\phantom{1}2.983&20.433&0.022&&&Mar 11&\phantom{1}7.064&20.482&0.023\\
Jan 12&\phantom{1}6.337&20.216&0.031&&&Jan 13&\phantom{1}7.931&20.110&0.013&&&Mar 11&\phantom{1}3.040&20.456&0.023&&&Mar 11&\phantom{1}7.112&20.458&0.023\\
Jan 12&\phantom{1}6.486&20.209&0.027&&&Jan 13&\phantom{1}8.005&20.126&0.012&&&Mar 11&\phantom{1}3.090&20.434&0.021&&&Mar 11&\phantom{1}7.161&20.465&0.022\\
Jan 12&\phantom{1}6.635&20.230&0.027&&&Jan 13&\phantom{1}8.079&20.136&0.013&&&Mar 11&\phantom{1}3.199&20.436&0.022&&&Mar 11&\phantom{1}7.209&20.446&0.021\\
Jan 12&\phantom{1}6.710&20.224&0.023&&&Jan 13&\phantom{1}8.153&20.114&0.013&&&Mar 11&\phantom{1}3.250&20.434&0.020&&&Mar 11&\phantom{1}7.258&20.455&0.021\\
Jan 12&\phantom{1}6.786&20.247&0.026&&&Jan 13&\phantom{1}8.226&20.119&0.013&&&Mar 11&\phantom{1}3.300&20.392&0.020&&&Mar 11&\phantom{1}7.307&20.444&0.022\\
Jan 12&\phantom{1}6.861&20.208&0.030&&&Jan 13&\phantom{1}8.301&20.144&0.013&&&Mar 11&\phantom{1}3.350&20.386&0.020&&&Mar 11&\phantom{1}7.355&20.449&0.023\\
Jan 12&\phantom{1}6.936&20.237&0.029&&&Jan 13&\phantom{1}8.374&20.152&0.012&&&Mar 11&\phantom{1}3.413&20.407&0.021&&&Mar 11&\phantom{1}7.403&20.478&0.025\\
Jan 12&\phantom{1}7.011&20.259&0.032&&&Jan 13&\phantom{1}8.448&20.177&0.013&&&Mar 11&\phantom{1}3.464&20.387&0.020&&&Mar 11&\phantom{1}7.451&20.455&0.023\\
Jan 12&\phantom{1}9.159&20.054&0.016&&&Jan 13&\phantom{1}8.522&20.170&0.013&&&Mar 11&\phantom{1}3.514&20.398&0.021&&&Mar 11&\phantom{1}7.503&20.446&0.021\\
Jan 12&\phantom{1}9.235&20.112&0.015&&&Jan 13&\phantom{1}8.597&20.173&0.013&&&Mar 11&\phantom{1}3.576&20.396&0.022&&&Mar 11&\phantom{1}7.552&20.459&0.021\\
Jan 12&\phantom{1}9.309&20.077&0.014&&&Jan 13&\phantom{1}8.676&20.166&0.012&&&Mar 11&\phantom{1}3.630&20.403&0.023&&&Mar 11&\phantom{1}7.600&20.431&0.020\\
Jan 12&\phantom{1}9.384&20.079&0.014&&&Jan 13&\phantom{1}8.750&20.180&0.012&&&Mar 11&\phantom{1}3.681&20.389&0.023&&&Mar 11&\phantom{1}7.648&20.417&0.019\\
Jan 12&\phantom{1}9.458&20.064&0.013&&&Jan 13&\phantom{1}8.824&20.169&0.013&&&Mar 11&\phantom{1}3.736&20.393&0.022&&&Mar 11&\phantom{1}7.696&20.421&0.020\\
Jan 12&\phantom{1}9.533&20.034&0.013&&&Jan 13&\phantom{1}8.898&20.182&0.013&&&Mar 11&\phantom{1}3.793&20.395&0.022&&&Mar 11&\phantom{1}7.746&20.429&0.022\\
Jan 12&\phantom{1}9.607&20.030&0.013&&&Jan 13&\phantom{1}9.195&20.213&0.013&&&Mar 11&\phantom{1}3.846&20.413&0.023&&&Mar 11&\phantom{1}7.795&20.435&0.022\\
Jan 12&\phantom{1}9.681&20.067&0.013&&&Jan 13&\phantom{1}9.268&20.220&0.013&&&Mar 11&\phantom{1}3.895&20.394&0.022&&&Mar 11&\phantom{1}7.843&20.406&0.022\\
Jan 12&\phantom{1}9.756&20.076&0.014&&&Jan 13&\phantom{1}9.342&20.224&0.014&&&Mar 11&\phantom{1}3.945&20.411&0.022&&&Mar 11&\phantom{1}7.891&20.461&0.023\\
Jan 12&\phantom{1}9.830&20.075&0.013&&&Jan 13&\phantom{1}9.416&20.213&0.013&&&Mar 11&\phantom{1}3.994&20.396&0.020&&&Mar 11&\phantom{1}7.939&20.445&0.023\\
Jan 12&\phantom{1}9.905&20.089&0.015&&&Jan 13&\phantom{1}9.490&20.216&0.013&&&Mar 11&\phantom{1}4.045&20.387&0.021&&&Mar 11&\phantom{1}7.990&20.418&0.022\\
Jan 12&\phantom{1}9.979&20.107&0.015&&&Jan 13&\phantom{1}9.564&20.235&0.013&&&Mar 11&\phantom{1}4.098&20.390&0.021&&&Mar 11&\phantom{1}8.039&20.443&0.021\\
Jan 12&10.055&20.116&0.014&&&Jan 13&\phantom{1}9.638&20.193&0.013&&&Mar 11&\phantom{1}4.147&20.415&0.022&&&Mar 11&\phantom{1}8.087&20.395&0.022\\
Jan 12&10.130&20.116&0.016&&&Jan 13&\phantom{1}9.934&20.176&0.013&&&Mar 11&\phantom{1}4.196&20.424&0.021&&&Mar 11&\phantom{1}8.135&20.431&0.020\\
Jan 12&10.428&20.114&0.022&&&Jan 13&10.008&20.181&0.012&&&Mar 11&\phantom{1}4.244&20.423&0.022&&&Mar 11&\phantom{1}8.184&20.418&0.020\\
Jan 12&10.576&20.136&0.023&&&Jan 13&10.082&20.167&0.013&&&Mar 11&\phantom{1}4.294&20.451&0.022&&&Mar 11&\phantom{1}8.237&20.430&0.021\\
Jan 12&10.651&20.132&0.016&&&Jan 13&10.156&20.172&0.012&&&Mar 11&\phantom{1}4.751&20.434&0.024&&&Mar 11&\phantom{1}8.285&20.391&0.021\\
Jan 12&10.725&20.128&0.017&&&Jan 13&10.230&20.126&0.012&&&Mar 11&\phantom{1}4.801&20.451&0.025&&&Mar 11&\phantom{1}8.334&20.443&0.022\\
Jan 12&10.800&20.139&0.016&&&Jan 13&10.304&20.132&0.012&&&Mar 11&\phantom{1}4.852&20.473&0.025&&&Mar 11&\phantom{1}8.382&20.400&0.022\\
Jan 12&10.873&20.110&0.017&&&Jan 13&10.380&20.133&0.012&&&Mar 11&\phantom{1}4.907&20.478&0.027&&&Mar 11&\phantom{1}8.430&20.422&0.024\\
Jan 12&10.948&20.106&0.016&&&Jan 13&10.454&20.136&0.012&&&Mar 11&\phantom{1}4.956&20.494&0.027&&&Mar 11&\phantom{1}8.496&20.394&0.023\\
Jan 12&11.022&20.136&0.015&&&Jan 13&10.528&20.150&0.012&&&Mar 11&\phantom{1}5.005&20.504&0.026&&&Mar 11&\phantom{1}8.545&20.407&0.023\\
Jan 12&11.096&20.159&0.015&&&Jan 13&10.601&20.119&0.012&&&Mar 11&\phantom{1}5.054&20.491&0.027&&&Mar 11&\phantom{1}8.593&20.440&0.024\\
Jan 12&11.170&20.131&0.014&&&Jan 13&10.675&20.104&0.012&&&Mar 11&\phantom{1}5.102&20.511&0.027&&&Mar 11&\phantom{1}8.641&20.419&0.024\\
Jan 12&11.245&20.114&0.014&&&Jan 13&10.750&20.103&0.012&&&Mar 11&\phantom{1}5.151&20.485&0.025&&&Mar 11&\phantom{1}8.689&20.464&0.025\\
Jan 12&11.319&20.106&0.014&&&Jan 13&10.824&20.088&0.012&&&Mar 11&\phantom{1}5.199&20.504&0.025&&&Mar 11&\phantom{1}8.837&20.473&0.026\\
Jan 12&11.394&20.158&0.015&&&Jan 13&10.898&20.067&0.012&&&Mar 11&\phantom{1}5.248&20.535&0.024&&&Mar 11&\phantom{1}8.886&20.468&0.026\\
Jan 12&11.544&20.145&0.023&&&Jan 13&10.972&20.070&0.012&&&Mar 11&\phantom{1}5.296&20.509&0.025&&&Mar 11&\phantom{1}8.935&20.472&0.027\\
Jan 12&11.916&20.149&0.023&&&Jan 13&11.046&20.081&0.012&&&Mar 11&\phantom{1}5.345&20.565&0.030&&&Mar 11&\phantom{1}8.983&20.488&0.027\\
Jan 12&12.065&20.111&0.023&&&Jan 13&11.121&20.063&0.012&&&Mar 11&\phantom{1}5.393&20.543&0.027&&&Mar 11&\phantom{1}9.031&20.455&0.028\\
Jan 12&12.215&20.077&0.016&&&Jan 13&11.417&20.056&0.012&&&Mar 11&\phantom{1}5.441&20.526&0.028&&&Mar 11&\phantom{1}9.079&20.499&0.030\\
Jan 12&12.363&20.119&0.014&&&Jan 13&11.490&20.038&0.012&&&Mar 11&\phantom{1}5.515&20.539&0.029&&&Mar 11&\phantom{1}9.129&20.447&0.032\\
Jan 12&12.688&20.045&0.016&&&Jan 13&11.565&20.048&0.012&&&Mar 11&\phantom{1}5.564&20.526&0.028&&&Mar 11&\phantom{1}9.178&20.494&0.032\\
Jan 13&\phantom{1}5.839&20.120&0.019&&&Jan 13&11.640&20.059&0.012&&&Mar 11&\phantom{1}5.612&20.508&0.023&&&Mar 11&\phantom{1}9.226&20.458&0.029\\
Jan 13&\phantom{1}5.987&20.140&0.015&&&Jan 13&11.713&20.051&0.012&&&Mar 11&\phantom{1}5.660&20.527&0.025&&&Mar 11&\phantom{1}9.274&20.501&0.031\\
Jan 13&\phantom{1}6.061&20.139&0.014&&&Jan 13&11.787&20.043&0.012&&&Mar 11&\phantom{1}5.708&20.520&0.025&&&Mar 11&\phantom{1}9.323&20.462&0.032\\
Jan 13&\phantom{1}6.136&20.150&0.014&&&Jan 13&11.861&20.057&0.012&&&Mar 11&\phantom{1}5.758&20.496&0.022&&&Mar 11&\phantom{1}9.373&20.494&0.034\\
Jan 13&\phantom{1}6.211&20.125&0.014&&&Jan 13&11.935&20.043&0.012&&&Mar 11&\phantom{1}5.806&20.497&0.024&&&Mar 11&\phantom{1}9.421&20.504&0.037\\
Jan 13&\phantom{1}6.285&20.119&0.014&&&Jan 13&12.009&20.062&0.013&&&Mar 11&\phantom{1}5.951&20.476&0.024&&&Mar 11&\phantom{1}9.469&20.536&0.038\\
Jan 13&\phantom{1}6.359&20.097&0.013&&&Jan 13&12.083&20.063&0.013&&&Mar 11&\phantom{1}6.002&20.541&0.024&&&Mar 11&\phantom{1}9.518&20.499&0.036\\
Jan 13&\phantom{1}6.433&20.105&0.013&&&Jan 13&12.157&20.051&0.013&&&Mar 11&\phantom{1}6.050&20.542&0.024&&&Mar 11&\phantom{1}9.566&20.537&0.038\\
Jan 13&\phantom{1}6.509&20.097&0.013&&&Jan 13&12.231&20.060&0.013&&&Mar 11&\phantom{1}6.099&20.529&0.025&&&Mar 11&\phantom{1}9.616&20.492&0.037\\
Jan 13&\phantom{1}6.583&20.100&0.013&&&Jan 13&12.305&20.071&0.013&&&Mar 11&\phantom{1}6.147&20.515&0.025&&&Mar 11&\phantom{1}9.664&20.489&0.043\\
Jan 13&\phantom{1}6.658&20.102&0.013&&&Jan 13&12.379&20.097&0.014&&&Mar 11&\phantom{1}6.195&20.530&0.027&&&Mar 11&\phantom{1}9.712&20.532&0.040\\
Jan 13&\phantom{1}6.731&20.112&0.013&&&Jan 13&12.454&20.093&0.014&&&Mar 11&\phantom{1}6.243&20.499&0.024&&&Mar 11&\phantom{1}9.760&20.596&0.042\\
Jan 13&\phantom{1}6.805&20.099&0.013&&&Jan 13&12.528&20.099&0.014&&&Mar 11&\phantom{1}6.292&20.552&0.024&&&Mar 11&\phantom{1}9.809&20.567&0.041\\
Jan 13&\phantom{1}6.879&20.089&0.013&&&Jan 13&12.602&20.096&0.015&&&Mar 11&\phantom{1}6.340&20.545&0.025&&&Mar 11&\phantom{1}9.866&20.607&0.042\\
Jan 13&\phantom{1}6.953&20.074&0.013&&&Jan 13&12.676&20.080&0.015&&&Mar 11&\phantom{1}6.388&20.475&0.024&&&Mar 11&\phantom{1}9.915&20.579&0.039\\
Jan 13&\phantom{1}7.027&20.099&0.013&&&Jan 13&12.750&20.106&0.015&&&Mar 11&\phantom{1}6.437&20.503&0.023&&&Mar 11&\phantom{1}9.963&20.599&0.041\\
Jan 13&\phantom{1}7.101&20.123&0.013&&&Jan 13&12.899&20.111&0.016&&&Mar 11&\phantom{1}6.675&20.486&0.038&&&Mar 11&10.011&20.533&0.040\\
Jan 13&\phantom{1}7.176&20.072&0.013&&&Jan 13&12.973&20.119&0.018&&&Mar 11&\phantom{1}6.722&20.506&0.025&&&Mar 11&10.059&20.609&0.044\\
Jan 13&\phantom{1}7.250&20.085&0.013&&&Jan 13&13.047&20.111&0.016&&&Mar 11&\phantom{1}6.773&20.473&0.023&&&Mar 11&10.109&20.537&0.046\\
Jan 13&\phantom{1}7.323&20.067&0.012&&&Jan 13&13.121&20.115&0.017&&&Mar 11&\phantom{1}6.821&20.462&0.023&&&Mar 11&10.157&20.592&0.048\\
Jan 13&\phantom{1}7.398&20.055&0.012&&&Jan 13&13.195&20.130&0.020&&&Mar 11&\phantom{1}6.869&20.478&0.024&&&Mar 11&10.206&20.582&0.049\\
Jan 13&\phantom{1}7.472&20.071&0.013&&&Jan 13&13.269&20.159&0.020&&&Mar 11&\phantom{1}6.917&20.479&0.022&&&Mar 11&10.254&20.610&0.053\\
Jan 13&\phantom{1}7.546&20.098&0.013&&&Jan 13&13.344&20.142&0.020&&&Mar 11&\phantom{1}6.966&20.460&0.023&&&Mar 11&10.302&20.561&0.051\\
Jan 13&\phantom{1}7.620&20.102&0.012&&&Mar 11&\phantom{1}2.850&20.446&0.033&&&Mar 11&\phantom{1}7.016&20.475&0.024&&&Mar 11&10.354&20.610&0.054\\


Apr 4&\phantom{1}3.028&20.766&0.044&&&Apr 4&\phantom{1}4.973&20.710&0.025&&&Apr 4&\phantom{1}7.177&20.829&0.031&&&Apr 6&\phantom{1}4.040&20.890&0.028\\
Apr 4&\phantom{1}3.071&20.767&0.027&&&Apr 4&\phantom{1}5.021&20.732&0.026&&&Apr 4&\phantom{1}7.225&20.853&0.031&&&Apr 6&\phantom{1}4.090&20.926&0.033\\
Apr 4&\phantom{1}3.127&20.775&0.028&&&Apr 4&\phantom{1}5.069&20.691&0.026&&&Apr 4&\phantom{1}7.275&20.795&0.031&&&Apr 6&\phantom{1}4.138&20.889&0.027\\
Apr 4&\phantom{1}3.176&20.773&0.028&&&Apr 4&\phantom{1}5.117&20.657&0.025&&&Apr 4&\phantom{1}7.323&20.842&0.032&&&Apr 6&\phantom{1}4.186&20.859&0.026\\
Apr 4&\phantom{1}3.224&20.794&0.028&&&Apr 4&\phantom{1}5.166&20.711&0.029&&&Apr 4&\phantom{1}7.371&20.794&0.033&&&Apr 6&\phantom{1}4.235&20.910&0.026\\
Apr 4&\phantom{1}3.275&20.723&0.028&&&Apr 4&\phantom{1}5.220&20.719&0.028&&&Apr 4&\phantom{1}7.419&20.844&0.033&&&Apr 6&\phantom{1}4.283&20.917&0.026\\
Apr 4&\phantom{1}3.323&20.706&0.025&&&Apr 4&\phantom{1}5.268&20.727&0.027&&&Apr 4&\phantom{1}7.468&20.833&0.033&&&Apr 6&\phantom{1}4.333&20.893&0.025\\
Apr 4&\phantom{1}3.371&20.738&0.027&&&Apr 4&\phantom{1}5.316&20.708&0.027&&&Apr 4&\phantom{1}7.522&20.802&0.034&&&Apr 6&\phantom{1}4.381&20.855&0.024\\
Apr 4&\phantom{1}3.420&20.776&0.028&&&Apr 4&\phantom{1}5.364&20.713&0.027&&&Apr 4&\phantom{1}7.570&20.830&0.036&&&Apr 6&\phantom{1}4.429&20.858&0.024\\
Apr 4&\phantom{1}3.468&20.758&0.029&&&Apr 4&\phantom{1}5.412&20.689&0.026&&&Apr 4&\phantom{1}7.618&20.820&0.037&&&Apr 6&\phantom{1}4.478&20.857&0.024\\
Apr 4&\phantom{1}3.517&20.722&0.028&&&Apr 4&\phantom{1}5.468&20.735&0.027&&&Apr 4&\phantom{1}7.666&20.745&0.037&&&Apr 6&\phantom{1}4.526&20.862&0.025\\
Apr 4&\phantom{1}3.566&20.772&0.027&&&Apr 4&\phantom{1}5.716&20.731&0.026&&&Apr 4&\phantom{1}7.715&20.824&0.042&&&Apr 6&\phantom{1}4.575&20.854&0.025\\
Apr 4&\phantom{1}3.614&20.724&0.025&&&Apr 4&\phantom{1}5.781&20.739&0.026&&&Apr 4&\phantom{1}7.764&20.766&0.042&&&Apr 6&\phantom{1}4.624&20.826&0.025\\
Apr 4&\phantom{1}3.662&20.764&0.027&&&Apr 4&\phantom{1}5.829&20.732&0.026&&&Apr 4&\phantom{1}7.812&20.797&0.037&&&Apr 6&\phantom{1}4.672&20.821&0.027\\
Apr 4&\phantom{1}3.711&20.759&0.026&&&Apr 4&\phantom{1}5.878&20.760&0.026&&&Apr 4&\phantom{1}7.860&20.804&0.036&&&Apr 6&\phantom{1}4.855&20.829&0.033\\
Apr 4&\phantom{1}3.760&20.769&0.026&&&Apr 4&\phantom{1}5.926&20.782&0.026&&&Apr 4&\phantom{1}7.908&20.835&0.038&&&Apr 6&\phantom{1}4.952&20.811&0.030\\
Apr 4&\phantom{1}3.808&20.735&0.026&&&Apr 4&\phantom{1}5.974&20.769&0.026&&&Apr 4&\phantom{1}7.957&20.814&0.038&&&Apr 6&\phantom{1}5.243&20.747&0.028\\
Apr 4&\phantom{1}3.857&20.750&0.025&&&Apr 4&\phantom{1}6.023&20.758&0.027&&&Apr 4&\phantom{1}8.006&20.855&0.037&&&Apr 6&\phantom{1}5.291&20.763&0.028\\
Apr 4&\phantom{1}3.905&20.728&0.026&&&Apr 4&\phantom{1}6.072&20.784&0.026&&&Apr 4&\phantom{1}8.054&20.839&0.038&&&Apr 6&\phantom{1}5.341&20.740&0.027\\
Apr 4&\phantom{1}3.953&20.784&0.027&&&Apr 4&\phantom{1}6.120&20.790&0.027&&&Apr 4&\phantom{1}8.102&20.773&0.034&&&Apr 6&\phantom{1}5.389&20.794&0.029\\
Apr 4&\phantom{1}4.002&20.714&0.028&&&Apr 4&\phantom{1}6.168&20.764&0.026&&&Apr 4&\phantom{1}8.151&20.781&0.035&&&Apr 6&\phantom{1}5.437&20.745&0.036\\
Apr 4&\phantom{1}4.051&20.756&0.028&&&Apr 4&\phantom{1}6.216&20.786&0.026&&&Apr 4&\phantom{1}8.199&20.797&0.037&&&Apr 6&\phantom{1}5.486&20.743&0.031\\
Apr 4&\phantom{1}4.099&20.758&0.026&&&Apr 4&\phantom{1}6.305&20.790&0.026&&&Apr 4&\phantom{1}8.250&20.737&0.036&&&Apr 6&\phantom{1}5.534&20.743&0.025\\
Apr 4&\phantom{1}4.147&20.710&0.025&&&Apr 4&\phantom{1}6.354&20.805&0.027&&&Apr 4&\phantom{1}8.298&20.763&0.038&&&Apr 6&\phantom{1}5.583&20.735&0.025\\
Apr 4&\phantom{1}4.195&20.724&0.026&&&Apr 4&\phantom{1}6.402&20.795&0.028&&&Apr 4&\phantom{1}8.346&20.741&0.038&&&Apr 6&\phantom{1}5.631&20.696&0.024\\
Apr 4&\phantom{1}4.244&20.744&0.028&&&Apr 4&\phantom{1}6.450&20.819&0.028&&&Apr 4&\phantom{1}8.394&20.750&0.053&&&Apr 6&\phantom{1}5.679&20.741&0.025\\
Apr 4&\phantom{1}4.292&20.710&0.025&&&Apr 4&\phantom{1}6.498&20.817&0.028&&&Apr 4&\phantom{1}8.442&20.754&0.045&&&Apr 6&\phantom{1}5.728&20.731&0.025\\
Apr 4&\phantom{1}4.341&20.719&0.024&&&Apr 4&\phantom{1}6.547&20.823&0.029&&&Apr 4&\phantom{1}8.498&20.798&0.051&&&Apr 6&\phantom{1}5.776&20.716&0.025\\
Apr 4&\phantom{1}4.389&20.716&0.024&&&Apr 4&\phantom{1}6.595&20.850&0.030&&&Apr 4&\phantom{1}8.546&20.756&0.048&&&Apr 6&\phantom{1}5.825&20.707&0.025\\
Apr 4&\phantom{1}4.438&20.684&0.026&&&Apr 4&\phantom{1}6.644&20.851&0.029&&&Apr 4&\phantom{1}8.594&20.751&0.049&&&Apr 6&\phantom{1}5.873&20.705&0.024\\
Apr 4&\phantom{1}4.488&20.698&0.023&&&Apr 4&\phantom{1}6.692&20.835&0.029&&&Apr 4&\phantom{1}8.642&20.699&0.050&&&Apr 6&\phantom{1}5.921&20.719&0.025\\
Apr 4&\phantom{1}4.536&20.687&0.023&&&Apr 4&\phantom{1}6.740&20.810&0.030&&&Apr 4&\phantom{1}8.740&20.729&0.055&&&Apr 6&\phantom{1}5.970&20.704&0.026\\
Apr 4&\phantom{1}4.584&20.712&0.023&&&Apr 4&\phantom{1}6.790&20.855&0.031&&&Apr 6&\phantom{1}3.029&20.856&0.134&&&Apr 6&\phantom{1}6.018&20.728&0.027\\
Apr 4&\phantom{1}4.632&20.724&0.023&&&Apr 4&\phantom{1}6.839&20.817&0.031&&&Apr 6&\phantom{1}3.646&20.858&0.027&&&Apr 6&\phantom{1}6.067&20.724&0.027\\
Apr 4&\phantom{1}4.680&20.719&0.023&&&Apr 4&\phantom{1}6.887&20.850&0.032&&&Apr 6&\phantom{1}3.701&20.865&0.027&&&Apr 6&\phantom{1}6.115&20.727&0.027\\
Apr 4&\phantom{1}4.730&20.727&0.024&&&Apr 4&\phantom{1}6.935&20.894&0.033&&&Apr 6&\phantom{1}3.749&20.920&0.027&&&Apr 6&\phantom{1}6.164&20.725&0.027\\
Apr 4&\phantom{1}4.779&20.699&0.025&&&Apr 4&\phantom{1}6.983&20.877&0.031&&&Apr 6&\phantom{1}3.797&20.875&0.027&&&Apr 6&\phantom{1}6.212&20.715&0.029\\
Apr 4&\phantom{1}4.827&20.672&0.026&&&Apr 4&\phantom{1}7.032&20.824&0.029&&&Apr 6&\phantom{1}3.847&20.888&0.025&&&Apr 6&\phantom{1}6.260&20.710&0.028\\
Apr 4&\phantom{1}4.875&20.684&0.025&&&Apr 4&\phantom{1}7.080&20.867&0.030&&&Apr 6&\phantom{1}3.895&20.901&0.027&&&\\
Apr 4&\phantom{1}4.923&20.711&0.025&&&Apr 4&\phantom{1}7.129&20.859&0.031&&&Apr 6&\phantom{1}3.992&20.901&0.028&&&\\


\hline
\enddata
\tablenotetext{a} {UT date of observations.}
\tablenotetext{b} {UT at midpoint of the exposure (uncorrected for light travel time).}
\tablenotetext{c} {Observed R-band magnitude (after applying absolute calibrations, extinction corrections, and comparison star corrections).}
\tablenotetext{d} {Uncertainty in the observed magnitude.}
\label{table2}
\end{deluxetable}


\begin{deluxetable}{lcccccc}  
\tabletypesize{\scriptsize}
\tablecolumns{7}  
\tablewidth{0pt} 
\setlength{\tabcolsep}{0.05in}
\tablecaption{Sidereal Rotation Periods and Lightcurve Amplitudes}
\tablehead{   
  \multicolumn{1}{l}{UT date\tablenotemark{a}}&
  \colhead{Synodic}&
  \multicolumn{2}{c}{Sidereal Period (hr)\tablenotemark{b}}&
  \colhead{Amplitude}&
  \colhead{Sub-Earth}&
  \colhead{Sub-Solar}\\
  \cmidrule(lr){3-4}
  \colhead{}&
  \colhead{Period (hr)}&
  \colhead{Prograde\tablenotemark{c}}&
  \colhead{Retrograde}&
  \colhead{(mag)}&
  \colhead{Latitude ($^\circ$)}&
  \colhead{Latitude ($^\circ$)}
}
\startdata
1987 Apr 1&---&---&---&0.3&$-$52&$-$46\\
1988 Apr 28&8.932 $\pm$ 0.001&8.931&8.935&0.5 -- 0.8&\phantom{1}$-$2 -- $-$10&$-$18 -- $+$17\\
1994 Nov 22&8.939 $\pm$ 0.003&8.939&8.941&0.5&\phantom{1}$-$3 -- \phantom{1}$+$4&\phantom{1}$+$2 -- \phantom{1}$-$7\\
1999 May 20&8.941 $\pm$ 0.002&8.939&8.945&0.5 -- 0.7&\phantom{1}$+$9 -- $+$10&\phantom{1}$-$1 -- $+$21\\
2010 Oct 25&8.950 $\pm$ 0.002&8.948&8.955&0.5 -- 0.8&$+$35 -- $+$43&$+$40 -- $+$13\\
2013 Feb 24&8.948 $\pm$ 0.001&8.948&8.950&0.2&$-$47&$-$48\\
\hline
\enddata
\tablenotetext{a} {Midpoint of the observations used to determine the synodic period. The relevant perihelion dates are 1988 Sep 16, 1994 Mar 16, 1999 Sep 8, 2005 Feb 15, 2010 Jul 4, and 2015 Nov 14.}
\tablenotetext{b} {The estimated uncertainties are identical to those listed in column 2.}
\tablenotetext{c} {The combined 2010 and 2013 results reveal that Tempel 2 is in prograde rotation.}
\label{table3}
\label{lasttable}
\end{deluxetable}


\renewcommand{\baselinestretch}{0.8}

\end{document}